**Electrical manipulation and detection of perpendicular altermagnetic order via proximitized Dirac semimetal**

Zhaohui Li[1,†], Wenqing He[1,†], Hua Bai[1,†], Yang Wang[2,†], Alexander J. Grutter[3], Guoyi Shi[1], Xiwen Zhang[2], Christy Kinane[4], Andrew Caruana[4], Hui Ru Tan[5], Yuchen Pu[1], Chenhui Zhang[1], Yongxi Wang[1], Hanbum Park[1], Anjan Soumyanarayanan[5,6], Lei Shen[2,7,*] & Hyunsoo Yang[1,*]

[1]Department of Electrical and Computer Engineering, National University of Singapore, Singapore 117583, Singapore.

[2]Department of Mechanical Engineering, National University of Singapore, Singapore 117575, Singapore.

[3]NIST Center for Neutron Research, National Institute of Standards and Technology, Gaithersburg, Maryland 20899-6102, USA.

[4]ISIS Facility, STFC Rutherford Appleton Laboratory, Harwell Science and Innovation Campus, Harwell, UK.

[5]Institute of Materials Research & Engineering, Agency for ScienceTechnology & Research (A*STAR), Singapore, Singapore.

[6]Department of Physics, National University of Singapore, Singapore, Singapore.

[7]National University of Singapore (Chong Qing) Research Institute; Chongqing Liang Jiang New Area, Chongqing 401123, China.

[*]Corresponding author. Email: shenlei@nus.edu.sg (L.S.); eleyang@nus.edu.sg (H.Y.)

[†]These authors contributed equally: Zhaohui Li, Wenqing He, Hua Bai, Yang Wang

**Altermagnets, which combine antiferromagnetic-like magnetic compensation with ferromagnetic-like broken time-reversal symmetry, hold great promise for high-density and ultrafast spintronic applications. However, the detection and switching of perpendicular altermagnetic order are fundamentally constrained by magnetic symmetry, restricting both fundamental studies and practical implementation. We realize robust electrical reading and deterministic switching of perpendicular altermagnetic order by designing a Dirac semimetal/altermagnet heterostructure of $PtTe_2$/CrSb. This engineered interface enables anomalous Hall readout via altermagnetic proximity effect and delivers efficient spin-orbit torque for manipulating the epitaxial perpendicular Néel vector in CrSb. These findings**

**significantly broaden the functional scope of altermagnetic heterostructures and pave the way for highly scalable altermagnetic memory.**

Electrical control of magnetic states is a cornerstone of both fundamental magnetism research and spintronic applications. Among various approaches, spin-orbit torque (SOT) has emerged as a powerful method for the electrical writing of magnetic information[1,2]. In ferromagnets, the switching of perpendicular magnetization offers significant advantages in terms of storage density, operation speed and energy efficiency[3-5]. In this scheme, the two stable magnetic states (up and down) represent binary data "0" and "1", making them ideal for memory bits. The anomalous Hall effect (AHE) is a prominent tool for macroscopic probing these perpendicular magnetic states. The non-zero anomalous Hall vector (**h**) is aligned with the magnetic moment (**m**) in ferromagnetic systems, leading to a Hall current described by $\mathbf{j}_{\mathrm{Hall}} = \mathbf{h} \times \mathbf{E}$, where **E** is the applied electric field[6]. This alignment (**h** // **m**) allows reliable readout of the two magnetic states in the perpendicular geometry via AHE. By contrast, while antiferromagnets exhibit appealing characteristics, including ultrafast spin dynamics and vanishing net moments[7,8], their intrinsic symmetry protection hinders conventional AHE-based readout.

The exploration of spin-split antiferromagnets unlocks new opportunities for antiferromagnet-based readout and writing[9-14]. Recent discoveries of altermagnets have propelled this field to a pivotal point[15-18], as they combine ferromagnetic-like non-relativistic spin splitting[19,20] with antiferromagnetic-like collinear magnetic compensation and high-frequency dynamics. These unique characteristics endow altermagnets with several advantages, including time-reversal symmetry breaking magnetic responses[21-26], negligible stray fields[27] and ultrafast operation speed[28], making them promising candidates for high-density and ultrafast magnetic random-access memory[29-31]. However, unlike ferromagnets, where **h** aligns naturally with the moments, the intricate magnetic symmetry of altermagnets typically confines **h** to directions orthogonal to the Néel vector (**L**), i.e., $\mathbf{h} \perp \mathbf{L}$ (ref. 32,33). Consequently, the spontaneous AHE can be only observed in altermagnets with in-plane **L** (ref. 21-23) (Fig. 1a), constraining device geometry and on-chip scalability, as SOT induced switching requires spin polarization perpendicular to **L**. This underscores the significance of detecting and switching perpendicular altermagnetic orders for both fundamental physics and potential device applications.

Recent theoretical studies establish altermagnetic heterostructures as a fertile platform for hosting a rich spectrum of magnetoelectric phenomena[34]. The concept of "altermagnetic proximity" has been proposed[34,35], in which the unique *k*-space spin splitting of altermagnets is imprinted onto the adjacent surface states. This not only enables the emergence of characteristics that are distinct or absent in the individual materials, but also allows the heterointerface to serve as a probe of the spin texture of altermagnets. Importantly, strong interfacial electron wavefunction interactions favor the realization of lower symmetries[36], thereby offering an alternative platform for inducing AHE in a perpendicular configuration.

Here we design a Dirac semimetal/altermagnet heterostructure composed of $PtTe_2$/CrSb. The functional topological/altermagnetic interface breaks these intrinsic symmetric constraints (Fig. 1a) and allows for an **h** along the Néel vector. We demonstrate theoretically and experimentally a large and spontaneous AHE, arising from the *PT*-symmetry breaking in CrSb and the induced band gap of interfacial orbitals via the altermagnetic proximity effect. Polarized neutron reflectometry (PNR) reveals a sizeable net moment of interfacial Pt atom, that is strongly coupled with the Néel vector, enabling reversal of the altermagnetic order, as evidenced by the exchange bias modulation in $PtTe_2$/CrSb/Fe trilayer and a CrSb-based magnetic tunneling junction (MTJ). Leveraging the efficient spin current from $PtTe_2$, we successfully achieve electrical manipulation of the perpendicular altermagnetic order in CrSb with high efficiency and robust endurance.

## $PtTe_2$/CrSb heterostructure

With a high Néel temperature and a large spin-splitting energy of ~ 1.2 eV, hexagonal CrSb, characterized by opposite-spin sublattices connected by sixfold rotation $C_{6z}$, has attracted significant interest as a *g*-wave altermagnet[37-39]. Although *PT*-symmetry breaking produces a finite momentum-dependent Berry curvature in bulk CrSb, its net contribution is completely canceled by the $\{C_{6z}T|00\frac{1}{2}\}$ symmetry operation, forbidding an anomalous Hall vector (Fig. S1)[32]. Topological/CrSb heterojunctions offer a unique approach to induce an altermagnetic order-dependent AHE. In particular, $PtTe_2$ as a Dirac semimetal exhibits topological surface states (Fig. S2) with a large spin Hall conductivity[40]. Moreover, with its unique electronic structure, Pt exhibits a density of states at the Fermi energy close to satisfying the Stoner criterion for ferromagnetism, making it a prototypical magnetic proximity material[41].

We therefore adopt the $PtTe_2$/CrSb heterostructure as a prototype and construct it using density functional theory to evaluate its band structure and Berry curvature (Fig. S3). The heterostructure reduces the magnetic space group to 156.51, lifting the $\{C_{6z}T|00\frac{1}{2}\}$-enforced cancellation of Berry curvature inherent to bulk CrSb and thereby yielding a finite Berry curvature, as illustrated in Fig. 1b. In addition, the strong interfacial interaction between Cr *d* orbitals and $PtTe_2$ opens band gaps near the Fermi level and creates new hotspots in momentum space, further enhancing the Berry curvature (Fig. 1c). Consequently, the $PtTe_2$/CrSb heterostructure exhibits a significantly larger anomalous Hall conductivity (AHC) than isolated altermagnets (Fig. 1d), with both the Berry curvature and AHC reversing sign upon Néel vector switching in CrSb (Fig. S4), enabling the readout of perpendicular altermagnetic order as shown in Fig. 1a. Furthermore, detailed analysis of the electronic states and control samples/models indicate that the non-zero Berry curvature primarily originates from the *d*-orbitals of altermagnetic CrSb, where *PT*-symmetry breaking in CrSb plays a decisive role (Figs. S5-8). For further details on the theoretical analysis of symmetry, Berry curvature, and electronic structure, see Supplementary Note I.

We prepare epitaxial heterostructures of a $PtTe_2$/CrSb bilayer on $Al_2O_3$(0001) substrates by molecular beam epitaxy (Supplementary Note II). The X-ray diffraction (Fig. S9) reveals that the CrSb layer has a characteristic (0001) texture. Fig. 2a shows interlayer antiferromagnetic coupling and the Néel vector along the *c* axis. The high crystal quality and epitaxial relationship of CrSb and $PtTe_2$ are further confirmed by the streaky reflection high-energy electron diffraction (RHEED) patterns (Fig. 2b and Fig. S10). Atomic force microscopy in Fig. S11 shows a smooth surface. Figure 2c and Fig. S12 illustrate the cross-sectional $(11\bar{2}0)$ plane and the atomically sharp interface of $PtTe_2$/CrSb via scanning transmission electron microscopy (STEM) image, attributed to the epitaxial interface between $PtTe_2$ and CrSb (~ 1% lattice mismatch).

We carry out temperature and angle-depending magneto-transport measurements. The resistivity $\rho_{xx}$ continuously increases with rising temperature (Fig. S13), representing typical metallic behavior. The dependence of longitudinal resistance ($R_{xx}$) on angle $\theta$ of $PtTe_2$ (5 nm)/CrSb (8 nm) is then measured at different temperatures, where the current is applied along the $x$ axis and the polarization of spin current is along the $y$ axis (Fig. 2d). A magnetic field of 5 T is applied in the plane perpendicular to the current and $\theta$ is the angle of magnetic field related to the $z$-axis. The observed magnetoresistance (MR) can be referred to as the spin Hall MR (SMR). A typical negative SMR can be observed at 300 K, aligning well with the antiferromagnetic phase[42]. At 390 K, however, the SMR becomes positive, indicating the CrSb film transitions to a paramagnetic state. We further investigate the SMR at temperatures ($T$) between 300 and 390 K (Fig. S14). The SMR ratio decreases progressively as $T$ increases and reaches zero at 350 K (Fig. 2e), which is considered as the Néel temperature ($T_N$). Above $T_N$, the SMR ratio transitions to positive and increases with $T$. Compared with bulk CrSb, the lower $T_N$ is attributed to the use of 8 nm thick CrSb layers. We also examine SMR for bilayers with varying thicknesses of CrSb (Supplementary Note III). A decrease of $T_N$ with reduced thickness is observed in Fig. S15. In addition, we characterize the antiferromagnetic behavior by exchange bias measurements in a $PtTe_2$/CrSb/Fe trilayer. The lateral negative shift of coercive fields, as shown in Fig. S16, results from the interfacial exchange bias. The blocking temperature is ~ 230 K, lower than the $T_N$ extracted by SMR measurements.

We turn towards the Hall measurements in the $PtTe_2$ (5 nm)/CrSb (8 nm) bilayer. Remarkably, despite being symmetry-forbidden in altermagnets with perpendicular magnetic anisotropy, a large spontaneous AHE is observed at both 10 K and 170 K, with remanence of approximately 100% (Fig. 2f). No AHE is observed in the control samples of CrSb and $PtTe_2$/Cr (Fig. S17), ruling out defects or doping under the growth conditions as the origin of the observed effect[43]. The AHE of $PtTe_2$/CrSb bilayer at different temperatures ($T$) exhibits clear hysteresis loops up to the Curie temperature ($T_C$) of 230 K (Fig. S18). A large AHC of around 73 S cm$^{-1}$ is estimated at 10 K, attributed to the breaking of $PT$ and $\{C_{6z}T\,|00\frac{1}{2}\}$ symmetries in CrSb and the orbital gap opening at the heterointerface, as revealed by the theoretical analysis above. In addition, the interfacial Pt atoms are expected to acquire a magnetic moment when adjacent to the CrSb with intralayer ferromagnetic coupling. The strong interaction between the interfacial moment and altermagnetic order produces a small Zeeman energy splitting and enables a detectable AHE. We emphasize that, although the interfacial moment is important for the observation of AHE, it does not constitute its origin. This is further corroborated by the extremely weak AHC in the $PtTe_2/Cr_2O_3$ sample (Fig. S6), despite the presence of interfacial moments[44,45]. Notably, the corresponding total magnetization $M_t$ of this sample in Fig. 2g shows a double-switching process around zero field and 460 mT. The red line indicates the interfacial magnetization $m_i$ at the $PtTe_2$/CrSb interface, exhibiting a consistent coercivity field with the AHE loop, which highlights the strong interfacial spin–Néel vector coupling. Moreover, the transition temperature of remanent magnetization is very close to that in $R_{xy}$ (Fig. S19). In contrast, the black line can be attributed to the net moment ($M_t - m_i$) from defects without contribution for $R_{xy}$ (ref. 22). Furthermore, as we vary the thickness of CrSb, the AHE remains robust in the bilayers (Figs. S20, S21 and Supplementary Note IV). The thickness dependence of $T_C$ and $T_N$ shows a similar pattern, indicating that the induced $m_i$ is strongly coupled with the altermagnetic order.

**Interfacial moment driven Néel vector switching**

We subsequently identify the interfacial moment driven Néel vector switching. To quantitatively investigate the depth-dependent net magnetization, we probe a representative $PtTe_2$/CrSb bilayer using polarized neutron reflectometry (PNR). Figure 3a shows the spin-dependent reflectivities alongside the best-fit nuclear (nSLD) and magnetic (mSLD) scattering length densities depth profiles along the film normal ($Q_Z$). To emphasize the magnetic features, Fig. 3b plots the spin asymmetry ratio, defined as the difference between the two non-spin-flip reflectivities normalized by their sum. We observe small but nonzero spin-splitting across a wide $Q$-range, indicating the presence of a net magnetization within the film. In particular, we note the existence of a large negative feature in the spin asymmetry above $Q_Z = 0.6$ nm$^{-1}$ at both 10 K and 100 K. In attempting to fit the data, we considered many physically plausible models of the magnetic depth profile, but only a model with a net magnetization at the top and bottom $PtTe_2$ interfaces can adequately capture this high-$Q$ splitting. In this model, the $PtTe_2$/CrSb interface possesses a magnetization of approximately 50 emu cm$^{-3}$ as shown in Fig. 3c, which is consistent with the magnetization measurements in Fig. 2g. Accounting for interface roughness, the magnetic moment at the $PtTe_2$/CrSb interface is estimated to be 0.4 $\mu_B$/Pt. While a net moment is also detected at the $Al_2O_3$/$PtTe_2$ interface, it originates in a region with significantly reduced nSLD, indicating a transitional growth region commonly observed in PNR measurements of layered tellurides[46]. While the origin of a net magnetization at this interface is unclear, we speculate it originates in impurities trapped in the more defective transitional growth region. This region most likely corresponds to the non-hysteretic switching feature shown in Fig. 2g and does not contribute to the AHE signal. At 10 K, the CrSb layer exhibits a vanishingly small net magnetization, approximately 6.4 ± 3.3 emu $cm^{-3}$, which is statistically indistinguishable from zero. See Figs. S22-26 and Supplementary Note V more details regarding the PNR fitting and statistical analysis.

The calculated electrostatic potential distribution along the *z*-axis of the heterostructure reveals a sharp discontinuity at the interface (Fig. S27 and Supplementary Note VI), indicating direct exchange and significant interfacial interactions, possibly because of the strong hybridization between the Cr 3*d* orbitals and the topological surface states of the Dirac semimetal $PtTe_2$ (ref. [47]). The spin-resolved projected density of states (PDOS) reveals nearly identical energy-dependent trends near the Fermi level of Cr, Te, and Pt atoms at the interface. This suggests strong orbital hybridization, further indicating the presence of a "Cr–Te–Pt" exchange interaction at the interface[48]. The robust interfacial bonding triggers the substantial charge redistribution at the interface, as evidenced by X-ray photoelectron spectroscopy (XPS) in Fig. S28, leading to significant reconstruction of orbital hybridization interacting adjacent Pt atoms. Given the unique density of states at the Fermi level of Pt akin to ferromagnetism, even slight alterations in its electronic structure can induce a $m_i$. DFT calculations for the $PtTe_2$/CrSb bilayer reveal that transferring 0.19 electrons/Pt at the interfacial monolayer results in a magnetic moment as large as 0.3 $\mu_B$/Pt.

The robust interfacial exchange interaction promotes reversal of the perpendicular altermagnetic order, ultimately giving rise to a measurable AHE[5,49]. To verify this conclusion, we conduct exchange-bias switching of a $PtTe_2$/CrSb/Fe trilayer. As shown in Fig. 3d, an exchange bias of 5 mT is detected via in-plane anisotropic magnetoresistance (AMR), originating from the interfacial pinning of the Fe magnetic moments by CrSb. After applying ±1 T out-of-plane magnetic fields, the in-plane exchange bias at the CrSb/Fe interface switches reproducibly between 5 and −5 mT, demonstrating reversal of the altermagnetic order in CrSb by the out-of-plane magnetic field[50]. We then fabricate a $PtTe_2$/CrSb/MgO/CoFeB MTJ and show that sweeping the magnetic field along the out-of-plane direction from 1 to −1 T and back to 1 T, yields a clear tunneling magnetoresistance (TMR) effect of ~ 0.3% (Fig. 2e). The characteristic field-dependent switching behavior matches the switching of both the CoFeB layer and the CrSb altermagnetic order (Fig. S29). The TMR originates from the spin-polarized current generated by CrSb due to its additional spin-splitting bands induced by strain[51] (see Figs. S30-33 and Supplementary Note VI for details). Therefore, all the theoretical and experimental results consistently demonstrate reversal of the perpendicular Néel vector in CrSb.

**Electrical switching of perpendicular altermagnetic order by SOT**

We examine the electrical switching of the perpendicular Néel vector in $PtTe_2$ (5 nm)/CrSb (8 nm) induced by SOT. The bottom layer of $PtTe_2$ with a large spin Hall conductivity serves as a spin source layer and induces a damping-like torque to precess the altermagnetic order. The interfacial magnetic moment-induced Zeeman energy breaks the switching symmetry of opposite altermagnetic states[5,49,52], thereby enabling deterministic switching of the Néel vector under an external magnetic field applied along different directions, as shown in Fig. 4a. As expected, Fig. 4b captures an apparent positive (negative) jump in the $R_{xy}$ under a negative (positive) current larger than a critical threshold ($I_c$) with an assistant magnetic field $\mu_0 H_x$ = 30 mT along the $x$ direction at 170 K. Opposite switching polarity can be observed when the direction of $\mu_0 H_x$ is reversed. The current density is calculated assuming a uniform current distribution across the bilayer. Moreover, we evaluate the heating effect caused by current injection at 170 K and 100 K. The real temperature of the sample during SOT switching is much below $T_C$ (Fig. S34 and Supplementary Note VII for details). These features suggest that SOT instead of thermal effects dominates the switching process of the Néel vector in CrSb.

We further study the magnetic field- and temperature-dependence of the SOT switching performance. With increasing the in-plane magnetic field $\mu_0 H_x$, the critical switching current $I_c$ remains almost constant (Figs. 4c and d), characteristic of antiferromagnetic order. This insensitivity arises because the magnetic field-induced Zeeman energy in antiferromagnets is much smaller than its anisotropic energy due to their negligible magnetization, which is a fundamental distinction from conventional ferromagnetic systems. In our heterostructure, the strong antiferromagnetic coupling makes altermagnetic order immune to the external magnetic field, which results in the independence of $I_c$ to $\mu_0 H_x$. These observations indicate switching of the altermagnetic order instead of only the interfacial moments, which is consistent with previous studies on antiferromagnets, such as $Mn_5Si_3$ (ref. 22), $Cr_2O_3$ (ref. 5) and $Mn_3Sn$ (refs. 4,10). Moreover, the insensitivity of $I_c$ to $\mu_0 H_x$ persists over a wide temperature range from 50 to 220 K (Fig. S35), and is also observed in devices with thinner CrSb layer (4 nm), where almost full switching is achieved as shown in Fig. S36. To examine the efficiency of electrical manipulation of altermagnetic order in CrSb, we turn to the dependence of the critical current density $J_c$ and $\mu_0 H_c$ at various temperatures. As the temperature decreases, $I_c$ linearly depends on $\mu_0 H_c$ as shown in Fig. 4e, which is consistent with previous studies on $Cr_2O_3$ (ref. 5). We calculate $\mu_0 H_c/J_c$ as a figure of merit to evaluate the SOT efficiency. It is found that $\mu_0 H_c/J_c$ is 11.8 mT cm$^2$ MA$^{-1}$, much larger than traditional ferromagnetic systems of ~ 1 mT cm$^2$ MA$^{-1}$ (refs. 1,2), indicating a higher efficiency in our heterostructure with perpendicular altermagnetic order.

**Theoretical calculations and macrospin model for altermagnetic order switching**

To explain the mechanism of field and SOT switching of the altermagnetic order, we employ DFT calculations, considering an A-type antiferromagnet (intralayer ferromagnetic and interlayer antiferromagnetic coupling) with perpendicular magnetic anisotropy. The large interfacial atomic magnetic moment and robust magnetic exchange coupling are crucial for altermagnetic order switching[5,49,52,53]. PNR and magnetization measurements consistently reveal a giant interfacial magnetization of 0.4 $\mu_B$/Pt. Based on these experimentally measured interfacial magnetic moments, DFT calculations demonstrate a significant interfacial magnetic coupling strength for the Cr–Te–Pt pathway, comparable in magnitude to the Cr–Cr super-exchange coupling within CrSb (Supplementary Note VIII)[47]. To further confirm the electrical manipulation of the altermagnetic order, we consider two cases: (1) $L$ switching and (2) only uncompensated $m_i$ at the interface is switched. By incorporating these parameters into an established heterostructure model, we calculate the energy states for both cases (Fig. S37). The results indicate that $L$ switching is energetically more favorable.

Furthermore, we construct a zero-temperature macrospin model with A-type antiferromagnetic exchange coupling $A_i$ and variable perpendicular magnetic anisotropies $K_i$ for different layers (see Supplementary Note IX for details). In this model, the first layer represents the magnetic moments induced by proximity effects, while the other layers correspond to the intrinsic magnetic moment within the bulk CrSb. Fig. S38 and S39 show field driven Néel vector switching and its layer-resolved switching dynamics, enabled by interfacial magnetic moments and exchange coupling, which reduces the energy barrier. SOT current induced switching can be expressed as $\boldsymbol{H}_{\mathrm{SOT}}^{i} = H_{\mathrm{SOT}}(\boldsymbol{m}_i \times \boldsymbol{\sigma})$, where $H_{\mathrm{SOT}}$ and $\boldsymbol{\sigma}$ denote the effective SOT magnetic field and the direction of spin polarization, respectively. The current-driven switching curves under both positive and negative $\mu_0 H_x$ are displayed in Fig. S40. These curves demonstrate that reversing external magnetic fields produces opposite switching polarities, aligning with our experimental findings in Fig. 4a. Importantly, the simulations reveal that in the absence of the $m_i$, SOT-induced deterministic switching cannot be achieved, thereby highlighting the significance of the induced $m_i$ in altermagnetic order switching. In addition, Fig. 4f depicts the magnetization-switching trajectories at high current densities under positive $\mu_0 H_x$. Due to the strong interlayer exchange interaction, the two spin sublattices remain nearly antiparallel and precess collectively in opposite directions during switching. This robust exchange interaction of the perpendicular Néel vector also results in high-frequency and simplified precession, enabling rapid switching dynamics (Fig. 4g). In addition, the simulations also show that the critical current is insensitive to the applied magnetic field, in agreement with our experiments. Our theoretical model provides a general insight into the mechanism of SOT-induced deterministic switching in CrSb[54].

Overall, the above analysis conclusively demonstrates that the $m_i$ plays a critical role not only in enabling the effective readout of the altermagnetic order but also in facilitating its deterministic 180° switching. The sophisticated interplay with interfacial spins reveals a promising strategy for achieving effective and rapid manipulation of altermagnetic order with perpendicular anisotropy. It is significant to mention that the unique advantage of altermagnets over conventional antiferromagnets lies in their momentum-locked spin splitting, which enables TMR dependent on longitudinal spin-polarized current[24,25], ultimately facilitating the realization of a practical perpendicular altermagnetic tunneling junction as shown in Fig. 3e (refs. 29,30). This makes the electrical switching of altermagnets important, especially for configurations with perpendicular magnetic anisotropy[3-5]. Furthermore, we measure the endurance for SOT switching by applying 6,000 alternating electrical pulses (Fig. S41 and Supplementary Note X). The Néel vector can be switched reversibly without obvious degradation, demonstrating its high robustness (Fig. 4h). These conditions lay the foundation for constructing reliable and robust SOT-based altermagnetic memory with high efficiency and reliability.

**Methods**:

**MBE growth of $PtTe_2$/CrSb heterostructures**

The $PtTe_2$/CrSb films were grown on sapphire $Al_2O_3$(0001) substrates using a molecular beam epitaxy (MBE) system with a base pressure of $5 \times 10^{-10}$ Torr. Prior to sample growth, pre-annealing processes were carried out at 600 °C for 30 min in the MBE chamber. During growth, the substrate temperature was maintained at 300 °C for $PtTe_2$, with a Te/Pt flux ratio exceeding 30 to prevent Te deficiency[40]. The growth rate of $PtTe_2$ was 0.1 nm $min^{-1}$. For CrSb growth on $PtTe_2$, the substrate temperature was 320 °C, the beam flux ratio of Sb:Cr was around 10:1 and the CrSb growth rate was 0.4 nm $min^{-1}$. For the $PtTe_2$/CrSb/Fe trilayer and $PtTe_2$/CrSb/MgO/CoFeB/Ta tunnel junction, the growth conditions of $PtTe_2$ and CrSb remained unchanged, while the Fe was in-situ deposited at room temperature in the MBE chamber, and the MgO/CoFeB/Ta was deposited by magnetron sputtering at room temperature with post annealing at 250 °C for 30 min. $Cr_2O_3$ film was grown by magnetron sputtering onto $Al_2O_3$(0001) substrates using magnetron sputtering. The deposition process was conducted using a $Cr_2O_3$ target at 400 °C. High-purity Cr, Te and Sb were evaporated from Knudsen cells, whereas Pt and Fe were evaporated from electron-beam evaporators. Reflection high-energy electron diffraction (RHEED) was used to monitor all the growth processes. Samples were then characterized by X-ray diffraction and atomic force microscope.

**Characterizations**

Magnetization measurements were carried out in a commercial superconducting quantum interference device magnetometer (MPMS3, Quantum Design). The magnetization loops as a function of the out-of-plane magnetic field from 10 K to 300 K were measured. For $PtTe_2$/CrSb/Fe trilayer, a 4 T field cooling process was performed first, followed by magnetic hysteresis (*M-H*) loop measurements under an in-plane magnetic field to determine the exchange bias field. The microstructures were examined using STEM (ARM-200CF, JEOL) at an accelerating voltage of 300 kV. XPS measurements of the Cr 2*p* and Te 3*d* core levels were carried out under high vacuum (~ $10^{-8}$ Torr).

**Polarized neutron reflectometry**

Polarized neutron reflectometry (PNR) experiments were performed on the Polref instrument at the ISIS Neutron and Muon Source. Polref is a white-beam time of flight (ToF) polarized neutron reflectometer utilizing a polarized wavelength band of 0.2 nm – 1.4 nm. Specimens were mounted horizontally in a helium flow cryostat with an exchange gas pressure of 4 kPa, providing a temperature stability of ±0.01K. Prior to measurement, the specimens underwent field-cooling from room temperature to 10 K while subjected to an in-plane magnetic field of 700 mT. Data collection employed a specular reflection setup, with the momentum transfer vector oriented normal to the film surface. Since magnetometry confirmed that the magnetization was saturated within the film plane under fields exceeding 700 mT, spin-flip scattering was deemed negligible and full polarization analysis was not required. Consequently, only the non-spin-flip components of the spin-resolved reflectivity were collected. The data were reduced using in-house Python code and analyzed using the Refl1D software package. Uncertainty estimation on fitted parameters was obtained using the DREAM Markov-Chain Monte Carlo algorithm implemented in the bumps python package, using $10^6$ samples.

**Device fabrication and transport measurements**

The films were patterned into 6 × 21 μm Hall bars using optical lithography combined with Ar ion milling. Following this, Ta/Cu/Ta electrodes were deposited onto the devices by magnetron sputtering. The transport characteristics were measured using a Hall bar with a physical property measurement system (PPMS, Quantum Design). For the SMR measurement, the longitudinal resistance ($R_{xx}$) was recorded as the devices were rotated in the *y*-*z* plane. The Hall resistance ($R_{xy}$) as a function of the applied magnetic field ($H_z$) was also measured to obtain the anomalous Hall resistance. In the switching measurements, a 10 ms pulse was initially applied, followed by a 500 ms wait time after the write current was turned off before measuring $R_{xy}$. A small reading current of 100 μA was then applied to measure the Hall resistance.

**Data and materials availability:** All data are available in the main text or the supplementary materials. The ISIS POLREF PNR data is available at the following DOI: https://doi.org/10.5286/ISIS.E.RB2400085-1.

**Code availability**: The codes that support the theoretical modelling of this study are available from the corresponding authors upon reasonable request.

**Acknowledgements:** This work is supported by the National Research Foundation (NRF) Singapore (NRF-CRP32-2024-0003, NRF-T-CRP-2025-0001), and Singapore MOE Tier 2 (MOE-T2EP50124-0006, T2EP50125-0008). Partial support for AJG was provided by the Center for High Resolution Neutron Scattering, a partnership between the National Institute of Standards and Technology and the National Science Foundation under Agreement No. DMR-2010792. Unless otherwise noted, NIST work was funded solely by the United States Government. Certain commercial equipment, instruments, software, or materials are identified in this paper in order to specify the experimental procedure adequately. Such identifications are not intended to imply recommendation or endorsement by NIST, nor it is intended to imply that the materials or equipment identified are necessarily the best available for the purpose.

**Author contributions:** H.Y., Z.L. and G.S. conceived and designed the experiments. Z.L. grew the samples. Z.L. carried out the reflection high-energy electron diffraction, atomic force microscopy, X-ray photoelectron spectroscopy, transport and switching measurements. Z.L., H.B. and W.H. analyzed the data. W.H. performed the macrospin simulation. A.J.G., C.J.K. and A.J.C. performed and analyzed the polarized neutron reflectometry measurements. C.Z. and H.P. carried out the X-ray diffraction measurements. Z.L., W.H. and Y.P. performed the device fabrications. H.P. contributed to the electrical measurements. Y.W. carried out the magnetization measurements. H.T. carried out the transmission electron microscopy measurements under the supervision of A.S.; Y. W. and X.Z. performed the first-principles calculations under the supervision of L.S.; Z.L., H.Y., H.B. and W.H. wrote the manuscript with contributions from all authors. H.Y. supervised the project. All authors discussed the results and commented on the manuscript.

**Conflict of Interests**: Authors declare that they have no competing interests.

## Figures and figure captions

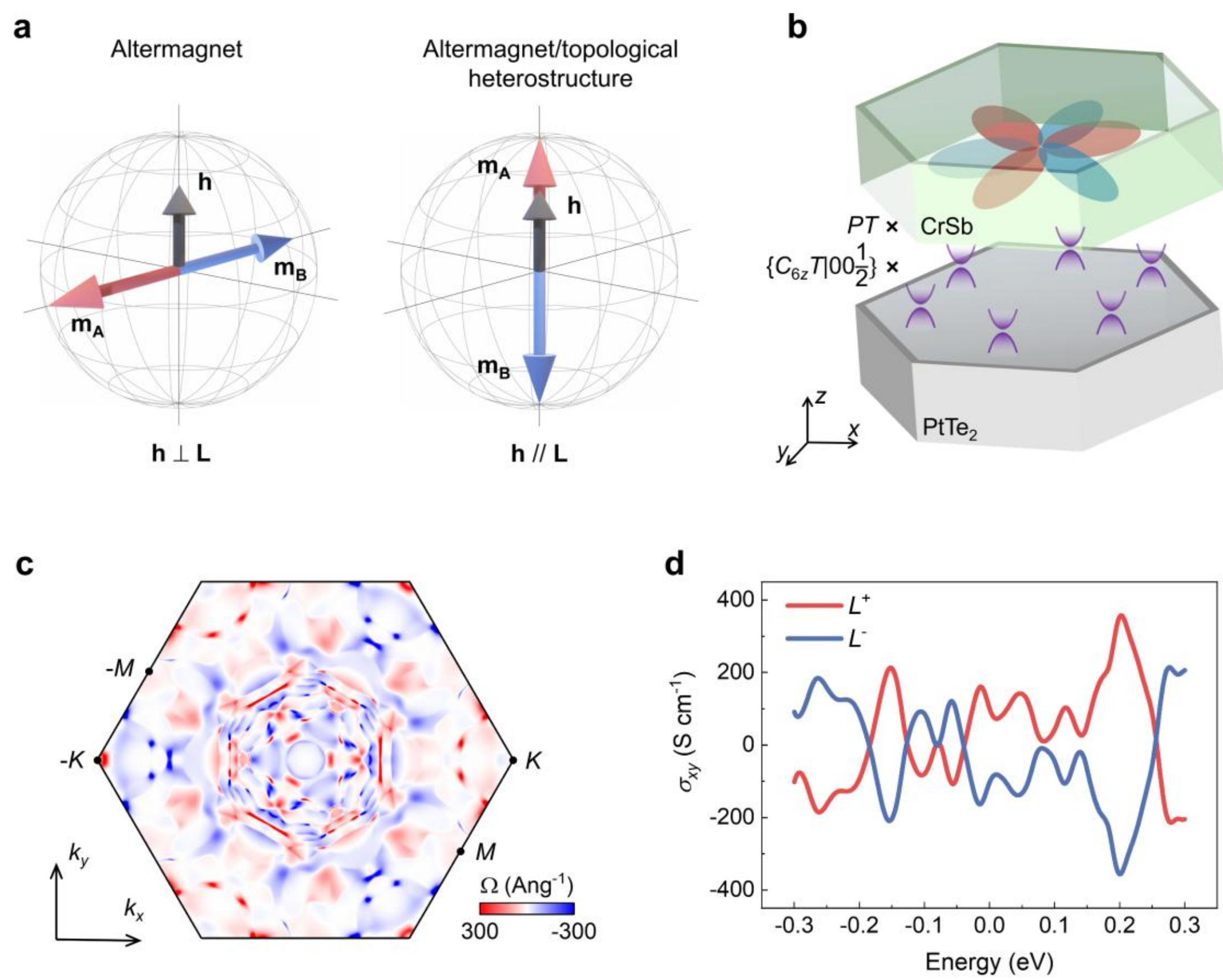

**Fig. 1. Anomalous Hall effect (AHE) in Dirac semimetal/altermagnet heterostructure of $PtTe_2$/CrSb with perpendicular magnetic anisotropy. a**, Comparison of altermagnet and altermagnet/topological heterostructure regarding the relationship between Néel vector (**L**) and Hall vector (**h**). The direction of the Néel vector is denoted by ($\mathbf{m_A} - \mathbf{m_B}$). The **h** is constrained to the direction perpendicular to the Néel vector in altermagnets (**h** ⊥ **L**), while out-of-plane **h** and **L** (**h** // **L**) in altermagnet/topological heterostructure are allowed. **b**, The anomalous Hall response in the $PtTe_2$/CrSb heterostructure is governed by the *PT*-symmetry breaking in CrSb and $\{C_{6z}T|00\frac{1}{2}\}$-symmetry breaking, with the interfacial hybridized orbital gaps further boosting its magnitude. **c**, Berry curvature near the Fermi level of the $PtTe_2$/CrSb heterostructure. **d**, Calculated $\sigma_{xy}$ near the Fermi level for the Néel vector along $L^+$ and $L^-$.

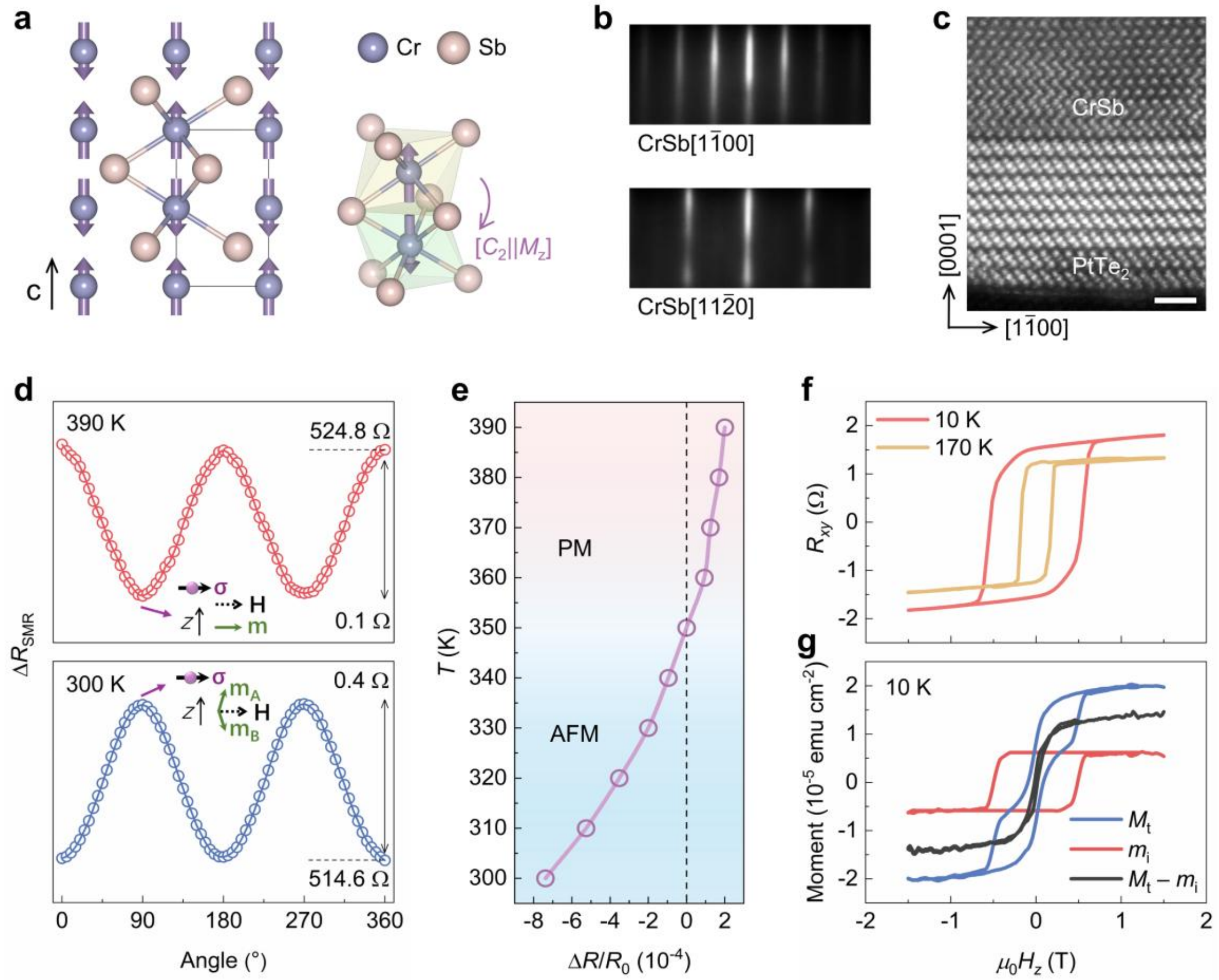

**Fig. 2. Crystal structure and magnetotransport of $PtTe_2$/CrSb bilayer. a**, Crystal structures and magnetic symmetry of CrSb. **b**, Reflection high-energy electron diffraction patterns along [11$\bar{2}$0] and [1$\bar{1}$00] directions of CrSb. **c**, Cross-sectional STEM image of the $PtTe_2$ (5 nm)/CrSb (8 nm) stack. The scale bar represents 1 nm. **d**, Representative angular dependencies of the longitudinal resistance variation ($\Delta R_{SMR}$) in the *y*-*z* plane under $\mu_0 H = 5$ T at 300 K and 390 K. We define the *x*-axis as the direction of the applied current, the *y*-axis as the direction perpendicular to *x* within film plane, the *z*-axis as the direction normal to the sample plane and the angle as the angle between the magnetic field and the *z*-axis. **e**, Temperature dependence of the SMR ratio from 300 K to 390 K. $\Delta R$ is defined as ($R$ (0°) – $R$ (90°)), while $R_0$ is defined as $R$ (0°). **f**, The anomalous Hall measurements at 10 K and 170 K. **g**, Corresponding magnetization hysteresis loops under an out-of-plane magnetic field ($\mu_0 H_z$) at 10 K. The total magnetization $M_t$ (blue line) can be decomposed into the interfacial magnetization $m_i$ (red line) and defect-induced one ($M_t - m_i$) (black line).

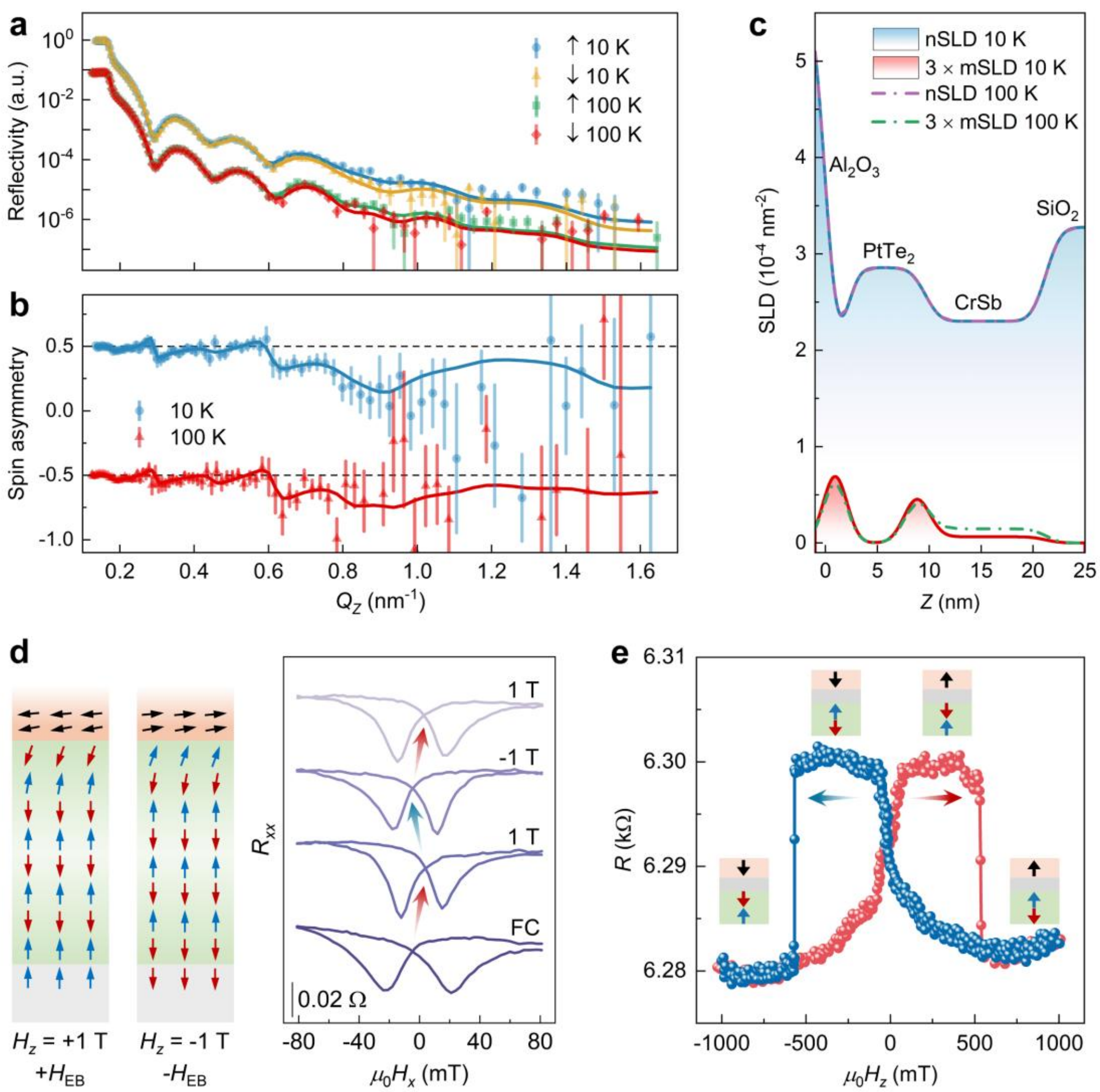

**Fig. 3. Spin textures and perpendicular Néel vector switching in $PtTe_2$/CrSb heterostructure. a**, Polarized neutron reflectivities with theoretical fits (at 10 K and 100 K in an in-plane field of 700 mT) of a $PtTe_2$/CrSb bilayer. **b**, The detailed spin asymmetry $(R^{++} - R^{--})/(R^{++} + R^{--})$ between the $R^{++}$ and $R^{--}$ channels, with theoretical fits. Offset for clarity, dashed lines indicating zero. Error bars represent ±1 standard deviation. **c**, The corresponding models with nSLD and mSLDs, used to obtain the fits shown, which contain magnetized $PtTe_2$ and CrSb layers for the heterostructure. **d**, The scheme of exchange bias manipulation of $PtTe_2$ (5 nm)/CrSb (8 nm)/Fe (8 nm) trilayer using ±1 T out-of-plane magnetic fields. The $PtTe_2$, CrSb, and Fe layers are shown in gray, green, and pink, respectively. In-plane anisotropic magnetoresistance (AMR) measurements at 10 K. A 4 T in-plane magnetic field was applied first for field cooling down to 10 K. A negative

exchange bias of −5 mT is observed in the AMR measurement. Subsequently, a +1 T out-of-plane magnetic field is applied at 10 K, and the in-plane AMR is recorded at 10 K. The procedure is repeated applying +1 T and −1 T out-of-plane field. **e**, Field dependence of tunnelling resistance measured in magnetic field at 10 K. The stack is $PtTe_2$ (8 nm)/CrSb (8 nm)/MgO (3 nm)/CoFeB (1.15 nm)/Ta (6 nm). Schematics for 4 switching states are shown in the inset. The CrSb, MgO, and CoFeB layers are shown in green, gray, and pink, respectively.

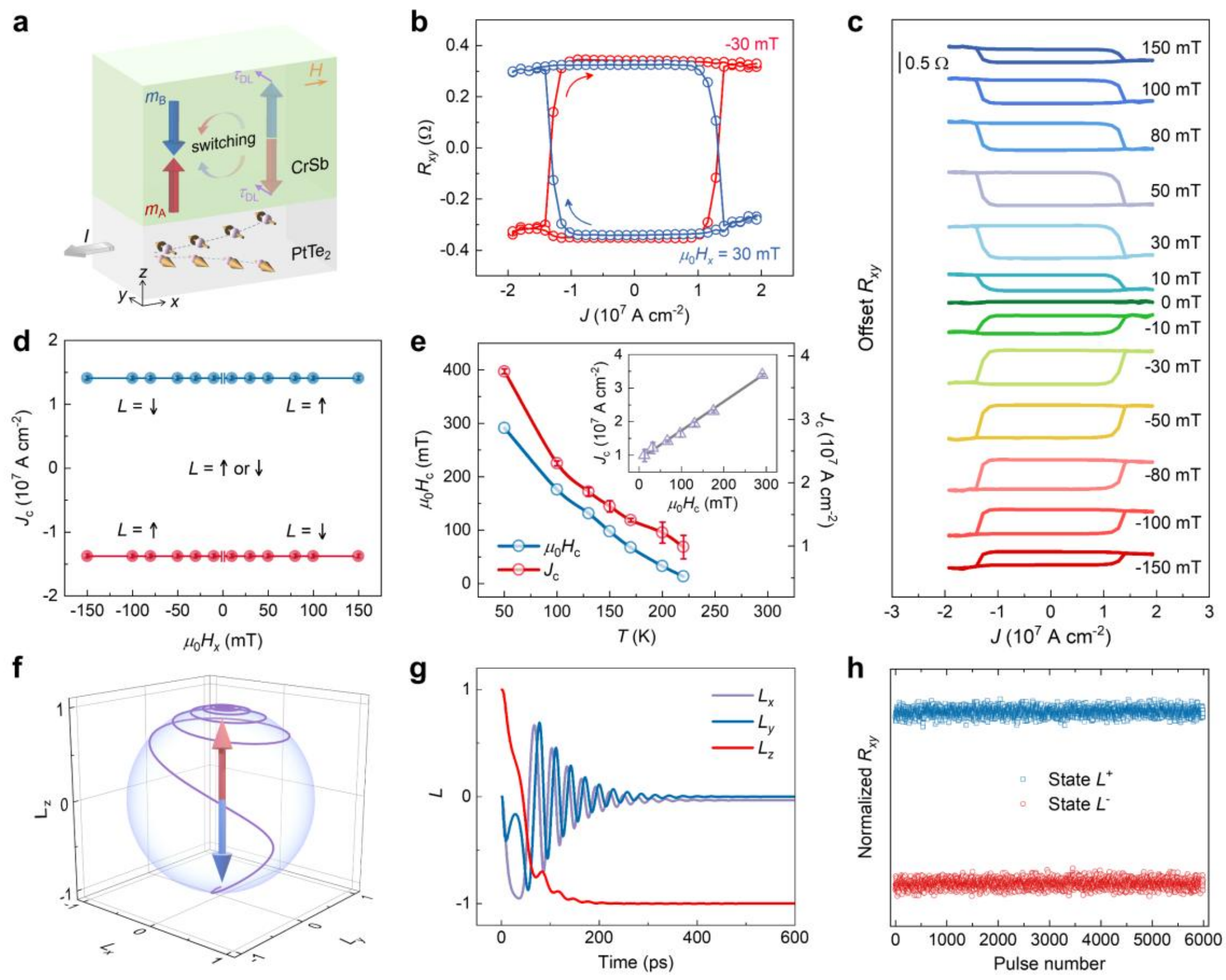

**Fig. 4. SOT-induced perpendicular altermagnetic order switching in $PtTe_2$/CrSb bilayers. a**, Schematics for SOT switching. The spin-polarized current generated in $PtTe_2$ exerts an SOT and causes switching of the altermagnetic order in CrSb under a pulse current and an external field along the $x$ direction. **b**, Hall resistance $R_{xy}$ as the function of the current $I$ of the $PtTe_2$ (5 nm)/CrSb (8 nm) device under opposite fields of ±30 mT at 170 K. **c**, $R_{xy} - I$ loops under different magnetic fields $\mu_0H_x$. **d**, The dependence of the critical switching current $I_c$ on $\mu_0H_x$. The error bars reflect the standard deviation of $R_{xy} - I$ loops under different fields. **e**, Temperature dependence of $\mu_0H_c$ and $I_c$. The corresponding dependence of $I_c$ on $\mu_0H_c$ is shown in the inset. The error bars reflect the standard deviation of measurements under different fields. **f**, Trajectories of Néel vector during the switching by macrospin simulations. **g**, Simulated dynamics of Néel vector in CrSb. **h**, Continuous cycling between the high $R_{xy}$ (state "$L^+$") and low $R_{xy}$ (state "$L^-$") by 180° switching of perpendicular Néel vector at 170 K.